# A Novel Heat Exchanger Design Method Using a Delayed Rejection Adaptive Metropolis Hasting Algorithm

Ahad Mohammadi[1], Javier Bonilla[2,3,*], Reza Zarghami[1,*], Shahab Golshan[1]

1: Process Design and Simulation Research Centre, School of Chemical Engineering, College of Engineering, University of Tehran, P.O. Box 11155/4563, Tehran, Iran

2: CIEMAT-PSA, Centro de Investigaciones Energéticas, Medioambientales y Tecnológicas, Plataforma Solar de Almería, Spain

3: CIESOL, Solar Energy Research Center, Joint Institute University of Almería - CIEMAT, Almería, Spain

**Abstract**

In this study, a shell-and-tube heat exchanger (STHX) design based on seven continuous independent design variables is proposed. Delayed Rejection Adaptive Metropolis hasting (DRAM) was utilized as a powerful tool in the Markov chain Monte Carlo (MCMC) sampling method. This Reverse Sampling (RS) method was used to find the probability distribution of design variables of the shell and tube heat exchanger. Thanks to this probability distribution, an uncertainty analysis was also performed to find the quality of these variables. In addition, a decision-making strategy based on confidence intervals of design variables and on the Total Annual Cost (TAC) provides the final selection of design variables. Results indicated high accuracies for the estimation of design variables which leads to marginally improved performance compared to commonly used optimization methods. In order to verify the capability of the proposed method, a case of study is also presented, it shows that a significant cost reduction is feasible with respect to multi-objective and single-objective optimization methods. Furthermore, the selected variables have good quality (in terms of probability distribution) and a lower TAC was also achieved. Results show that the costs of the proposed design are lower than those obtained from optimization method reported in previous studies. The algorithm was also used to determine the impact of using probability values for the design variables rather than single values to obtain the best heat transfer area and pumping power. In particular, a reduction of the TAC up to 3.5% was achieved in the case considered.

---

[*] Corresponding authors:
Tel.: +34 950 387 800; fax: +34 950 365 015; E-mail: javier.bonilla@psa.es (J. Bonilla).
Tel.: +98 21 6696 7797; fax: +98 21 6646 1024; E-mail: rzarghami@ut.ac.ir (R. Zarghami).

**1. Introduction**

Heat exchangers are significant and integral components in chemical industries, they are used for a variety of applications including energy saving, exchange, and recovery[1]. Shell and tube heat exchangers (STHXs) are commonly used in chemical processes, power plants, and air conditioning, thanks to the numerous advantages they offer over other types of heat exchangers[2]. An efficient heat exchanger contributes to lower the consumption of the energy resources and materials, providing both economic and environmental benefits. The most conventional method used for heat exchanger design is the iterative process based on trial and error. This approach highly relies on the designer's experience and generally leads to over-designed parameters [3]. Several shell and tube heat exchanger design methods are discussed in handbooks, which are generally based on trial-and-error approaches [2, 4-6].

Optimization of heat exchangers has been studied by different researchers where optimization algorithms have been used to find the optimum design parameters (the optimization algorithm is selected based on the type of variables: discrete or continuous) [3]. Wildi-Tremblay and Gosselin [7] presented a procedure for minimizing the cost of a shell-and-tube heat exchanger based on genetic algorithms. In their work, evaluations of the performances of heat exchangers were based on an adapted version of the Bell–Delaware method [8]. Their results showed that the procedure can properly and rapidly identify the optimal design for a specified heat transfer process.

Selbaset et al. [9] proposed the application of a genetic algorithm (GA) for optimal design of shell-and-tube heat exchangers. In their research, approximate design methods were investigated and a generalized procedure was developed to run the GA algorithm in order to find the global minimum heat exchanger area. The authors found out that combinatorial algorithms, such as genetic algorithms, provide significant improvement compared to traditional design strategies for finding the optimum design. Hadidi et al.[10] developed a new shell and tube heat exchanger optimization design approach based on a biogeography-based optimization (BBO) algorithm. They applied the BBO technique to minimize the total cost of the equipment including capital investment and the sum of discounted annual energy expenditures related to pumping in the heat exchanger. Their results indicated that the BBO algorithm could be successfully applied for the design of shell and tube heat exchangers.

Ponce-Ortega et al. [11] used a genetic algorithm for the optimal design of shell-and-tube heat exchangers. They used Bell–Delaware correlations [8] to properly calculate heat transfer coefficients and pressure drops in the shell-side, considering the minimization of the Total Annual Cost (TAC) as an objective function. Fesanghary et al.[12] used a harmony search algorithm for minimizing the total cost of shell and tube heat exchangers. They applied global sensitivity analysis to identify the geometrical parameters that have the largest impact on the total cost of STHXs. Their results revealed that the proposed algorithm could converge to optimal solutions with higher accuracy than genetic algorithm.

Caputo et al. [13] carried out economic optimizations of heat exchanger designs using GAs. They proposed a method for the design of shell and tube heat exchangers based on a genetic algorithm. They achieved a reduction of capital investment up to 7.4% and savings in operating costs up to 93%, with an overall reduction of total cost up to 52%.

Hadidi [14] investigated a robust approach for optimal design of plate fin heat exchangers using a biogeography based optimization algorithm. The author's parametric analysis was carried out to evaluate the sensitivity of the proposed method with respect to the cost and structural parameters. Özçelik [15] developed and applied a genetic algorithm to estimate the optimal values of discrete and continuous variables in Mixed Integer Non Linear Programming (MINLP) test problems. Their results over the test problems showed that the programmed algorithm could estimate acceptable values of continuous variables and optimal values of integer variables. Finally, such algorithm was extended for parametric studies and for finding optimum configuration of heat exchangers.

Hilbert et al. [16] developed a multi-objective optimization approach based on a genetic algorithm to find the most favourable geometry to simultaneously maximize the blade shape of the heat exchanger while at the same time minimizing the pressure loss. They considered the coupling of flow / heat transfer processes. Turgut [17] proposed a hybrid approach, entitled

Hybrid Chaotic Quantum behaved Particle Swarm Optimization (HCQPSO), for thermal design of plate fin heat exchangers. He tested the algorithm efficiency with different benchmark problems and compared them with those of other metaheuristic algorithms. His results revealed that HCQPSO finds far better solutions minimizing the objective functions compared to designs and methods reported in the literature.

Ayala et al.[18] presented a Multi-Objective Free Search approach combined with Differential Evolution (MOFSDE) for heat exchanger optimization. Their results indicated that MOFSDE shows better performance than the Non-dominated Sorting Genetic AlgorithmII (NSGA-II).

Huang et al. [19] proposed a multi-objective design optimization strategy based on genetic algorithms for U-tube vertical heat exchangers. Their results showed that the proposed strategy can decrease the total cost of the system (i.e. the upfront cost and 20 years' of operation cost) by 9.5% as compared to the original design. Compared to a single-objective design optimization strategy, 6.2% more energy could be saved by using their multi-objective design optimization strategy.Habimana [20] developed a model using NSGA-II for the design optimization together with the MCMC method for uncertainty analysis. His model minimized the area of the system and the momentary heat recovery output.

Results of optimization algorithms and traditional design methods are expressed as single values for design variables which do not give any information about the quality and uncertainty of the parameters.

In the present work, the values of the target variables (heat exchanger area and pump power) were estimated by sampling design variables using the DRAM method to obtain the heat exchanger design variables distributions. The design variables are: baffle spacing, baffle cut, tube-to-baffle diametrical clearance, shell-to-baffle diametrical clearance, tube length, tube outer diameter, and tube wall thickness, which were estimated considering their uncertainty bands rather than a fixed value. With this algorithm, the presented uncertainty analysis could be accomplished more thoroughly and uncertainty bands could be studied more accurate than in comparison with other methods. In addition, a conventional shell and tube heat exchanger, as a case of study, was used for the validation of the proposed technique and results were compared to multi-objective and single-objective optimization methods. Finally, a cost function is defined and a decision making process was established for the selection of the design variables based on their estimated distributions and TAC.

## 2. Sampling Methods

Sampling can be used to predict the behaviour of a particular model under a set of defined circumstances in order to find appropriate values for the model parameters by fitting model results to experimental data [21]. One of the most significant benefits of using sampling algorithms is the ability of these methods to analyse uncertainties. In sampling methods, the decision-making process is essential for the selection of the final variables among a set of samples.

In statistics, Markov Chain Monte Carlo (MCMC) methods are algorithms for sampling: Adaptive Metropolis (AM) and Delayed Rejection (DR) are two methods for improving the MCMC performance [22]. The main insight behind AM is its ability to perform on-line tuning of the proposed distribution based on the past sample path of the chain [23]. The rationale behind adaptive strategies is to learn from the information obtained during the run of the chain, and based on this, to efficiently tune the proposals. The acceptance probability of the second stage candidate is computed so that the reversibility of the Markov chain relative to the intuition distribution is preserved. The basic idea of DR is based upond rejection of a proposed candidate point, instead of retaining the same position, a second stage move is proposed [24].The details of the DRAM method are presented in appendix 2.

In this paper, DRAM was used to obtain the probability distribution of design variables of the shell and tube heat exchanger. An advantage of this method is its ability to express the design variables in the form of probabilistic distributions with a confidence interval. Confidence intervals provide an essential understanding of how much faith we can have in our sampling and provide the most likely range for the unknown population of all variables.Unlike optimization algorithms, sampling methods can represent the uncertainty of variables. Uncertainty analysis is useful in real processes where deviations from set points affect the performance of the system. To perform a quantitative uncertainty analysis, probability distributions should be assigned to each design variable.

*2.1. Design Variables*

Seven continuous decision variables are considered in the sampling process in order to obtain the values of the area and pumping power based on the design algorithm presented in Appendix 1. The DRAM method changes the values of these seven variables. The algorithm in Appendix 1 takes those values and calculates the area and pumping power. This sampling procedure run until stable probability distributions of the design variables, heat exchanger area and pumping power, are obtained. The probability distributions of the design variables are assumed to be Gaussian. The specifications of the design variables are as follows [3]:

- The baffle spacing at the centre, inlet, and outlet ($L_{bc} = L_{bo} = L_{bi}$) varies between the minimum baffle spacing of 0.0508 m and the maximum supported tube span of 0.2540 m (referring to $X_1$ in fig.1) [25].

- The baffle cut ($B_c$) can vary from 15 % to 45 % ( referring to $X_2$ in fig.1).

- Tube-to-baffle diameter clearance ($\delta_{tb}$) can take values between $0.01 d_o$ m and $0.1 d_o$ m (referring to $X_3$ in fig.1).

- Shell-to-baffle diametrical clearance ($\delta_{sb}$) is between 0.0032 m and 0.011 *m* [26] (referring to $X_4$ in fig.1).

- The tube length (*L*) : 2.438 and 11.58 *m* [26] (referring to $X_5$ in fig.1).

- The tube outer diameter ($d_o$): [0.01588 to 0.0508] *m* (referring to $X_6$ in fig.1).

- The tube wall thickness: [1.651, 4.572] *mm* (referring to $X_7$ in fig.1).

A schematic sampling network is illustrated in Fig. 1, which corresponds with the third step of the process represented in Fig. 2.

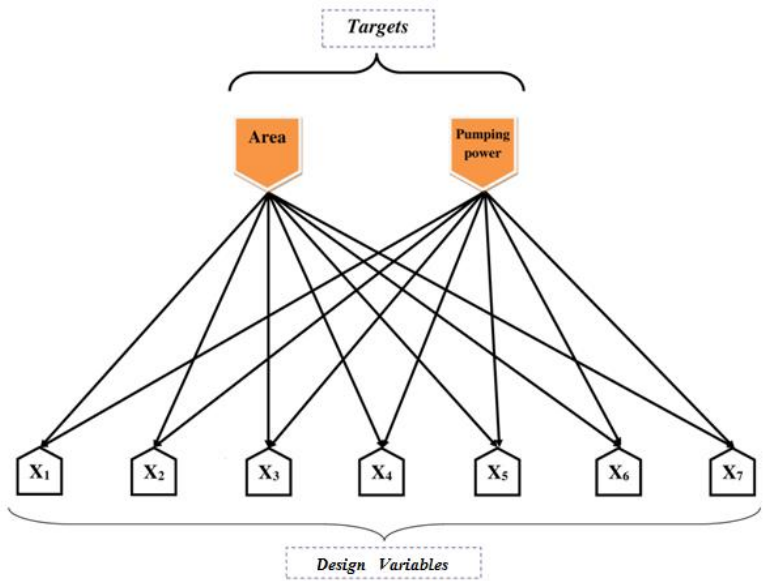

Fig. 1. Schematics of design and target variables.

The top nodes are target variables (area and pumping power) and the bottom nodes are design variables. In this figure, the target variables dependence on the design variables is shown.

## 3. Results and Discussion

DRAM was carried out for a case of study selected from the literature [7] to test the performance of the sampling algorithm for obtaining values for the design variables of a shell-and-tube heat exchanger. In addition, the results of the proposed method were compared with previous methods for design optimization [3]. For this case of study, 30,000 samples were considered.

Fig. 2 shows the designing procedure flowchart, which is proposed and used in this research. Seven steps can be found on the flowchart which are described in the following:

- Step (1): The governing equations on the shell-and-tube heat exchanger, design variables and target variables are specified.
- Step (2): The initial values of the design variables, used in the DRAM method, are set according to their ranges.
- Step (3): The DRAM algorithm performs the sampling of the design variables in this step.
- Step (4): The heat exchanger area and pumping power are calculated for each instance of the design variables in this step.
- Step (5): The new calculated values are compared with the present values in order to reach the stable distribution of the design variables.
  (The probability distribution of design variables does not change when increasing number of samples. As the number of samples increases, the probability distribution of the design variables converges to a unique distribution. It became clear that when the number of samples exceeds 30,000, a stable distribution is obtained.)
- Step (6): Samples are obtained for each design variable. (When the sampling is completed, for each one of the design variables, a Gaussian probability distribution with given variance and mean values is obtained.)
- Step (7): The selection of a set of the design variables, based on the decision making strategy (see Fig. 4), is performed in this step.

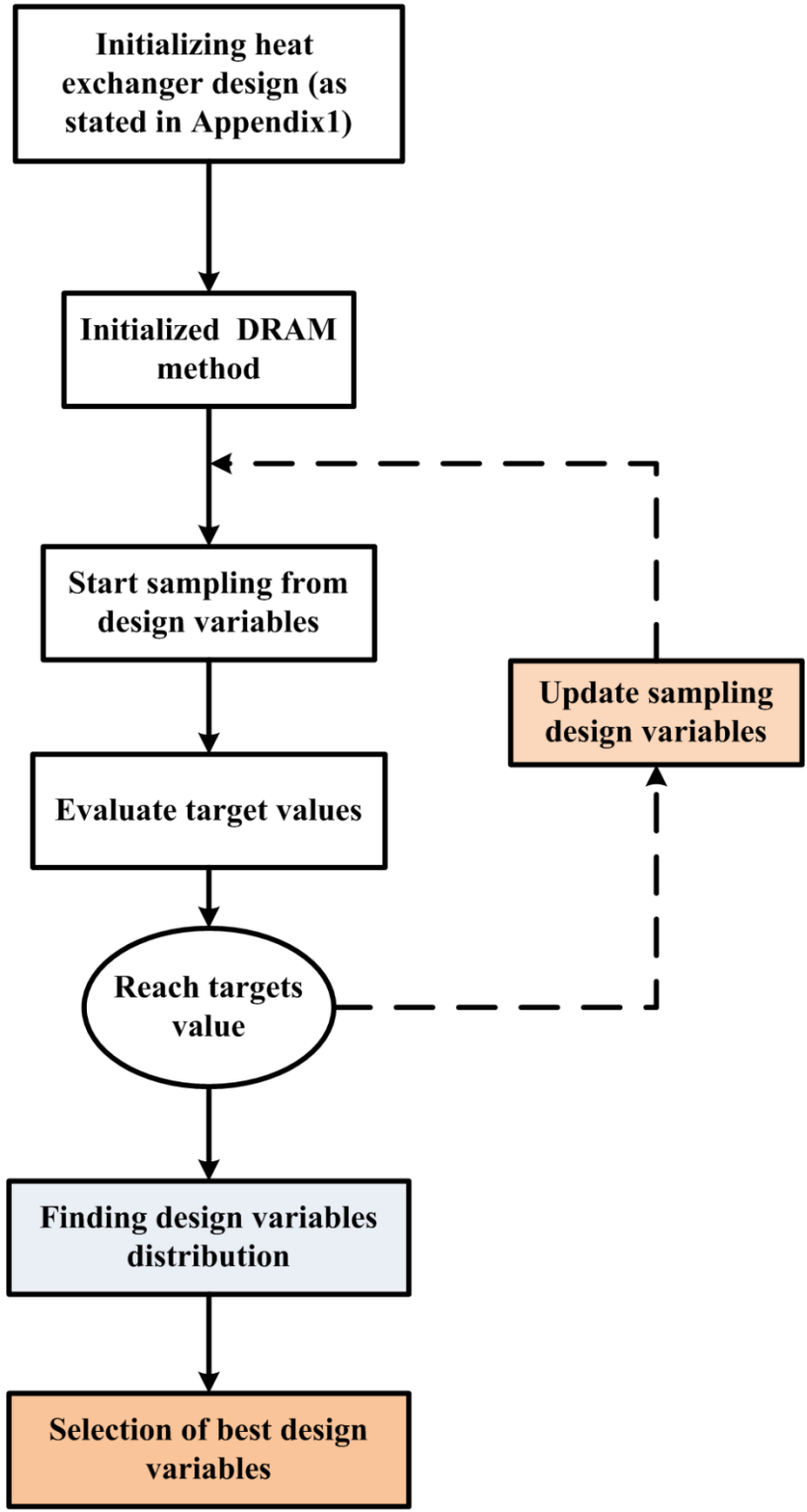

Fig. 2. Flowchart of the design procedure

For the initial guess, the sampling begins with the arithmetic mean value of the design variables in their ranges, and a variance value of ±2σ of the limit for each variable. It should be considered that according to the DRAM method; every random choice for the mean value in this limit makes the sampling algorithm converge [27]. Similarly, for the specific variance, 99.8 % of data are located inside -3σ to +3σ (σ denotes variance); however in this work initial limits are set to ±2σ of the limit as stated before, in order to have a higher confidence. The design variables are improved towards the desired target variable values at each sampling iteration. The sampling process continues up to the point where all the nodes, including the design variables, reach their stable distributions.

A confidence interval of 90% was considered for the design variables. Fig. 3 depicts the samples of the design variables and the resulting overall heat exchanger area for each set of samples. Two elliptical shapes are shown in this figure; the inner and outer curves illustrate the regions that contain 50% and 90 % of the samples around their mean values, respectively.

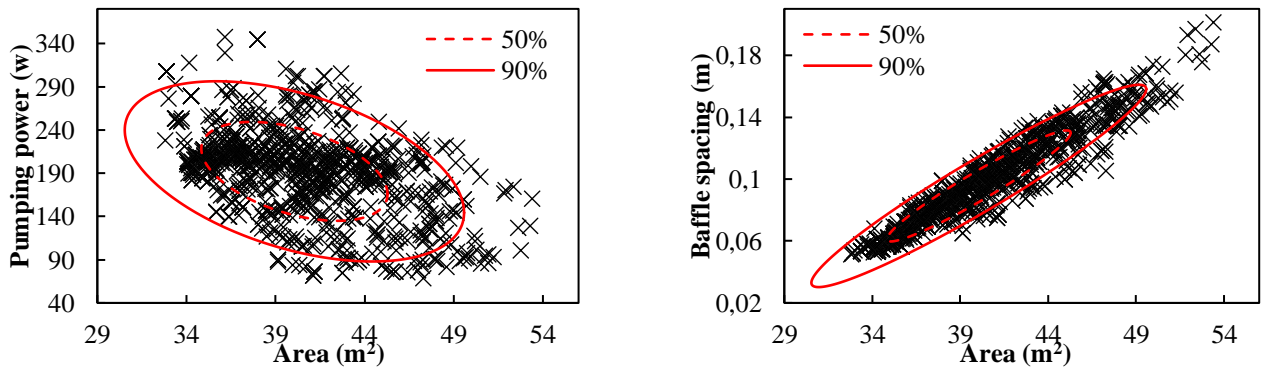

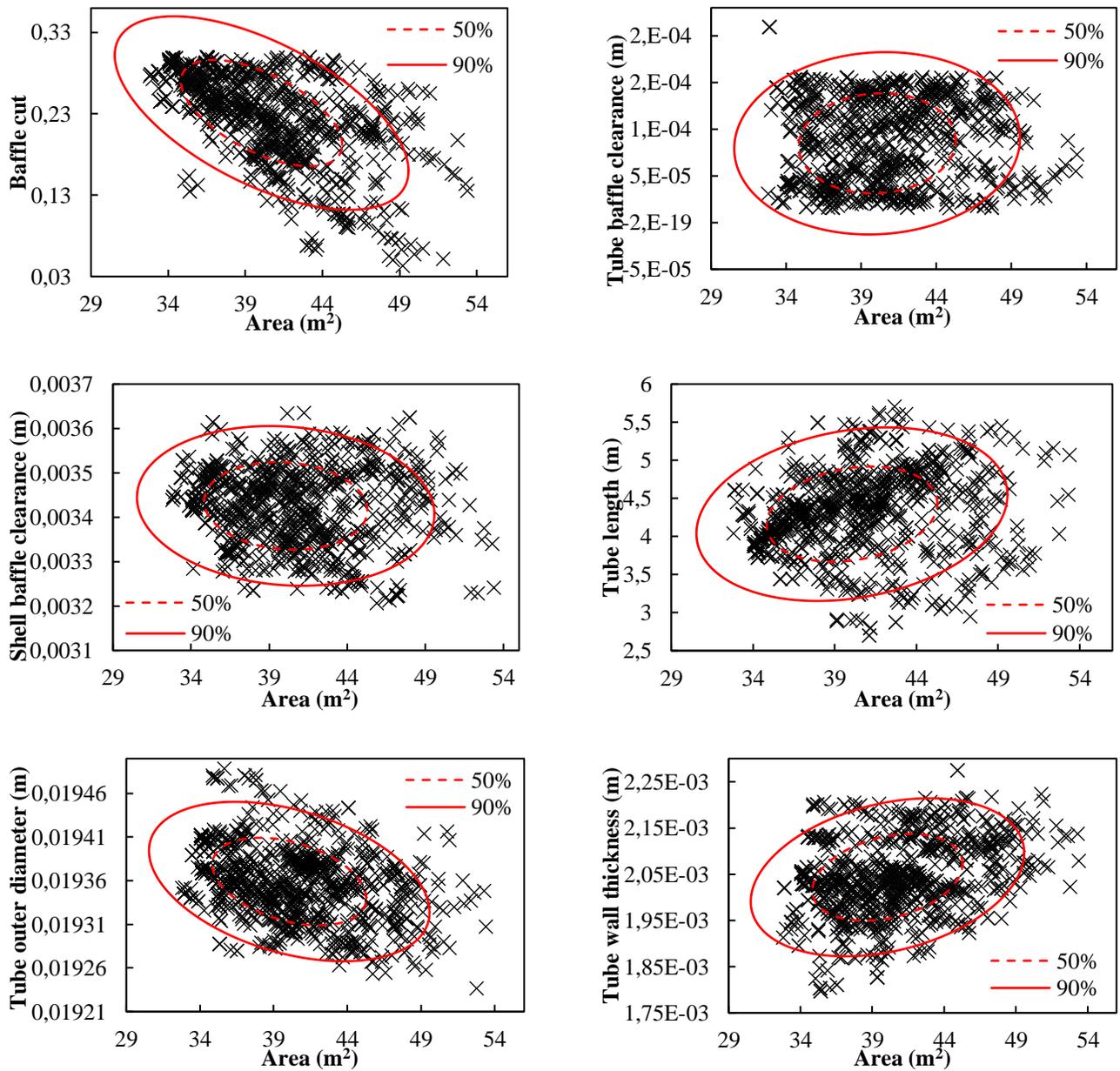

Fig. 3. Probability distribution of design variables as a function of the heat exchange surface area

In order to test the modelling results, a case of study considering the cooling of Naphtha using water in a shell-and-tube heat exchanger is presented. The operational data is used for validation (see Table 1). The results from the sampling method, in terms of mean and variance of the Gaussian distributions of the design variables, have been compared to results in the literature [7] in Table 2. It should be noted that most of the approaches reported in literature for heat exchanger optimization are based on single values for the design variables, but in this study, we present probability distributions for all the design variables.

|  | Tube-side | Shell-side |
|---|---|---|
| **Fluid** | Cooling water | Naphtha |
| **Flow rate (kg/s)** | 30 | 2.7 |
| **Inlet temperature (°C)** | 33 | 114 |
| **Outlet temperature (°C)** | 37.21 | 40 |
| **Density (kg/m$^3$)** | 1000 | 656 |
| **Heat capacity (J/(kg K))** | 4186.8 | 2646.06 |
| **Viscosity (N s/m$^2$)** | 0.00071 | $3.70 * 10^{-4}$ |
| **Thermal conductivity (W/m K)** | 0.63 | 0.11 |
| **Design pressure (Pa)** | 1278142 | 738767 |
| **Fouling resistance (m$^2$ K/W)** | 0.0004 | 0.0002 |
| **Material of construction** | Stainless steel | Carbon steel |
| **Wall thermal conductivity (W/(m K))** | 16 | 55 |

Table 1. Design data used for validation [7]

| **Parameter** | DRAM | | **Single-objective optimization** | **Multi-objective optimization** |
|---|---|---|---|---|
| | mean | Variance | | |
| **Baffle spacing(m)** | 0.0956 | 9.2e-03 | 0.06 | 0.079 |
| **Baffle cut (%)** | 23.10 | 3.1 | 25 | 16.515 |
| **Tube-to-baffle diameter clearance(mm)** | 0.24864 | 9.0e-2 | 0.381 | 0.204 |
| **Shell-to-baffle diametrical clearance(mm)** | 3.4 | 7.0e-01 | 3 | 3.279 |
| **The tube length(m)** | 4.292 | 0.881 | 10.7 | 3.426 |
| **The tube outer diameter(mm)** | 23.4 | 3.813 | 38.1 | 19.578 |
| **The tube wall thickness(mm)** | 2.05 | 0.634 | 3.405 | 1.652 |

Table 2. DRAM calculated variables compared to multi-objective and single-objective optimizations

The optimum values from [3] are used as target function values for the sampling algorithm. The sampling method performed 30,000 iterations. The results show good performance with respect to the design variables estimation compared to the optimization methods. As estimated, incrementing the number of iterations leads to more accurate results. By increasing the number of samples, it was found that the result values almost did not change and it can be stated that convergence was achieved, which indicates a stable distribution of the design variables.

Fig. 4 presents the mean values and uncertainty bands of the independent variables used in the DRAM sampling compared to the results from multi-objective and single-objective optimizations. As stated previously, the results of optimization algorithms are described as fixed values without any estimation about the possible deviations from these points. As illustrated in Fig.4, the uncertainty band of the tube to baffle diameter clearance has the widest band among all variables, while the tube outer diameter, tube wall thickness, tube length and baffle spacing variables have narrow distributions around their mean values. In other words, these variables are less likely to deviate from their mean. Additionally, this probability distribution can help in the decision-making process. If, under any circumstance, the manufacturer cannot use the proposed design values, then other values for the design variables can be selected.

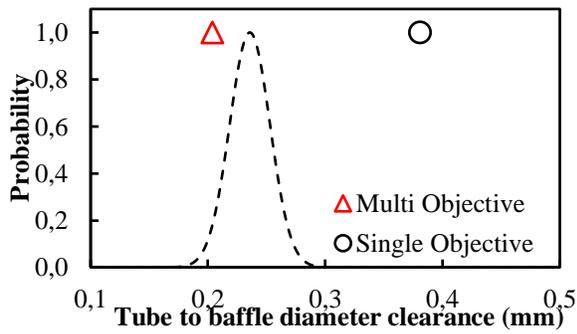
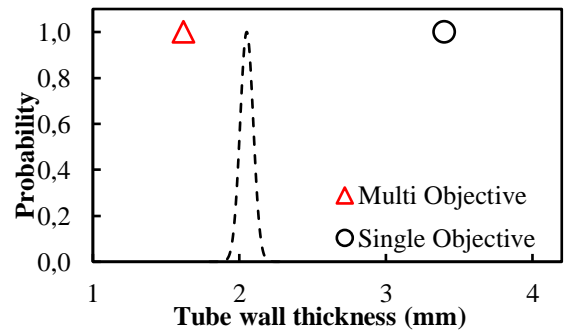
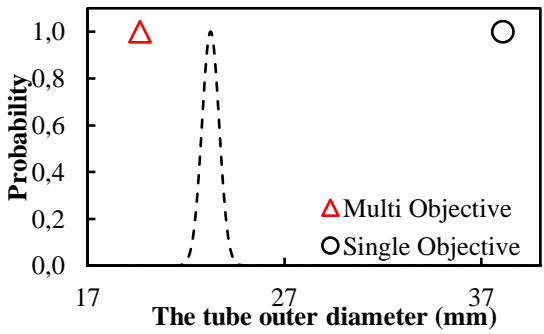
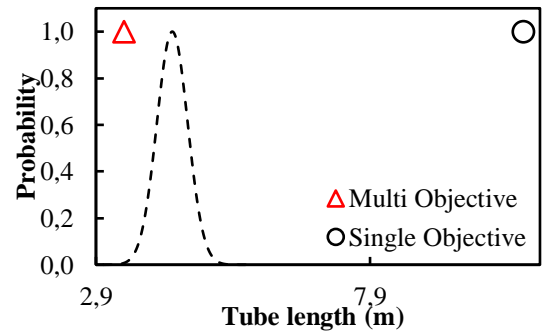
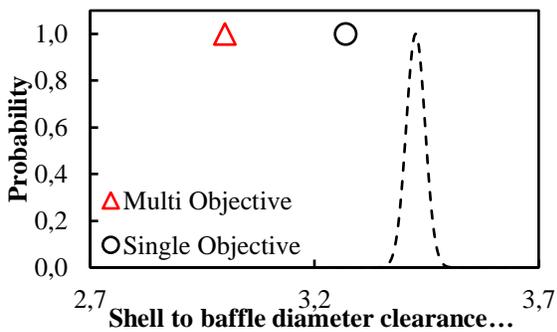
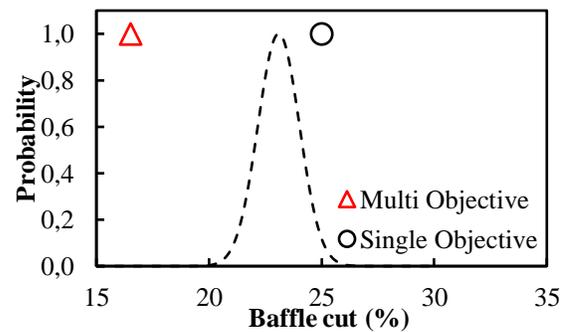

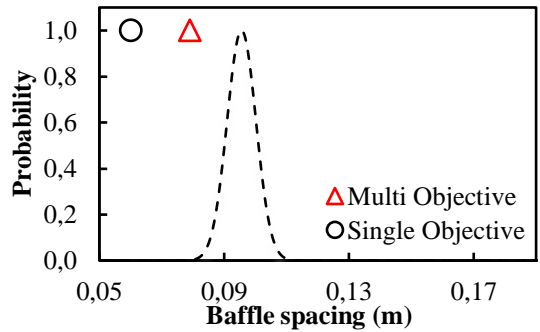

Fig. 4. Probability distribution of design variables compared to multi-objective and single-objective optimizations

A cost analysis method was used for the selection of the final design variables in the decision-making process. The global annual cost includes the operating cost and the initial cost expressed in terms of annuities, it was used in the decision making process for selecting the values of the design variables.

Fig. 5 sketches the decision making process, which involves the following steps:

- Goal: To get the best values for the design variables.
- Step (1): Select TAC as the decision variable.
- Step (2): Obtain sets of design variables with the DRAM method and the multi-objective genetic algorithm.
- Step (3): Calculate the TAC for all sets of design variables that are obtained from both methods.
- Step (4) Choose the set of design variables which has the lowest TAC.

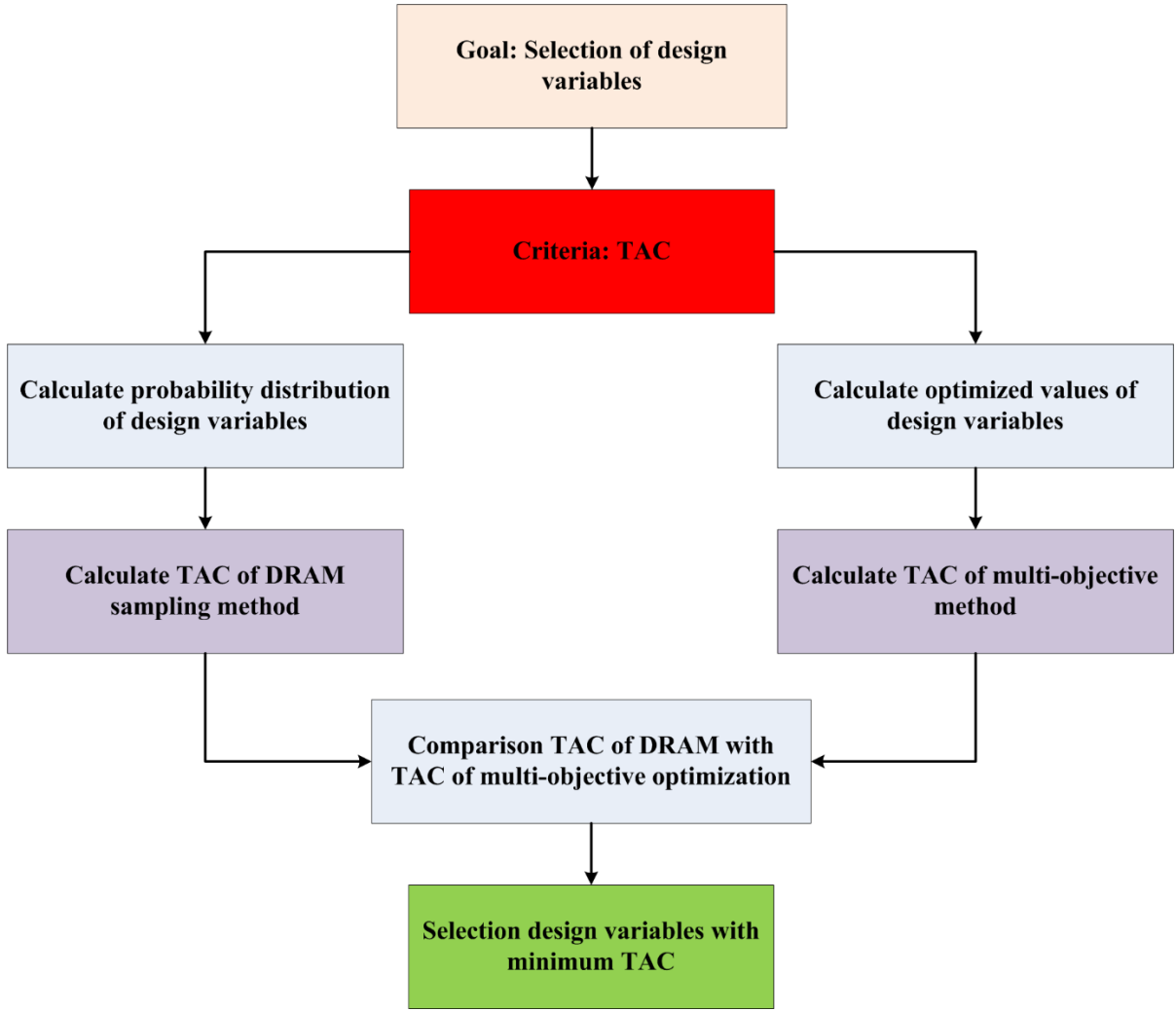

Fig 5. Flowchart of the decision making process

Fig. 6 shows the TAC associated with each sample together with the obtained from the multi-objective optimization. In Fig.6, a lower TAC means better design variables. The sets of design variables predicted by the DRAM method have lower TAC compared to the multi-objective optimization in most of the cases.

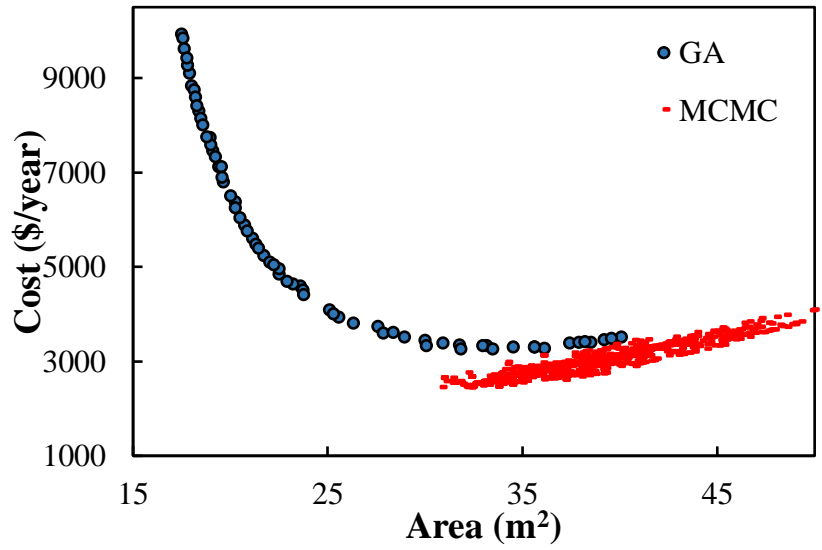

Fig. 6. Cost analysis for decision-making of multi-objective optimization [3] vs. DRAM sampling

From the results presented in Table 3, it can be inferred that the DRAM algorithm is suitable for the design of shell and tube heat exchangers. In the studied case, the TAC was drastically reduced and a significant percent of TAC reduction was achieved compared to the traditional design method. A TAC percentage decrease of 3.52% was obtained, which confirms the effectiveness of the proposed approach. As a final remark, it should be noted that the sampling method design generally leads to a heat exchanger structure markedly different from that considered as a conventional design.

| Parameter | DRAM | | Single-objective Optimization | Multi-objective Optimization |
|---|---|---|---|---|
| | Mean | variance | | |
| $A_o$ (m$^2$) | 37.16 | 5.32 | 37.14 | 37.14 |
| $\Delta P_S$, shell-side (Pa) | 21632 | 1543.3 | 22600 | 20620 |

| ΔP$_t$ tube side (Pa) | 8697 | 1223 | 8600 | 8584 |

Table 3. Comparison between sampling algorithm, multi-objective and single-objective optimizations using 30,000 samples

## 4. Conclusions

The DRAM sampling method was used to design a shell-and-tube heat exchanger. Based on the proposed approach, an algorithm was developed, and a case of study was investigated to illustrate its effectiveness, efficiency and suitability of the proposed method. Uncertainty analysis through uncertainty bands measured, and the deviations from mean values, were plotted for all of the design parameters. A decision-making strategy based on a confidence interval and TAC was used for final design variable selection. Results demonstrate accurate design variables values when compared with those obtained from optimization methods. The DRAM method presents lower TAC by choosing more appropriate design variable values.

Results indicate that DRAM can provide an estimation of the confidence interval for design variables, comparing the interval range to previously published values.. This method shows that design variables, in addition to the most likely value, have a confidence interval for evaluating their quality. The main advantages of the sampling method are that the designer can choose one particular solution from distribution solutions based on the proposed desicion making process, as well as estimating the cost of a particular design. With respect to the test case, a reduction of TCA of 3.52% was obtained, which shows the potential of the proposed method for improving the design of more cost-effective heat exchangers.

**Acknowledgements**

The results presented in this paper have been obtained within collaboration with IPS Company. The financial support is gratefully acknowledged.

**Appendix 1**

Fig. 7 shows a schematic view of a shell-and-tube heat exchanger. The overall heat transfer coefficient based on tube outer diameter $U_0$ is given by (1) [2].

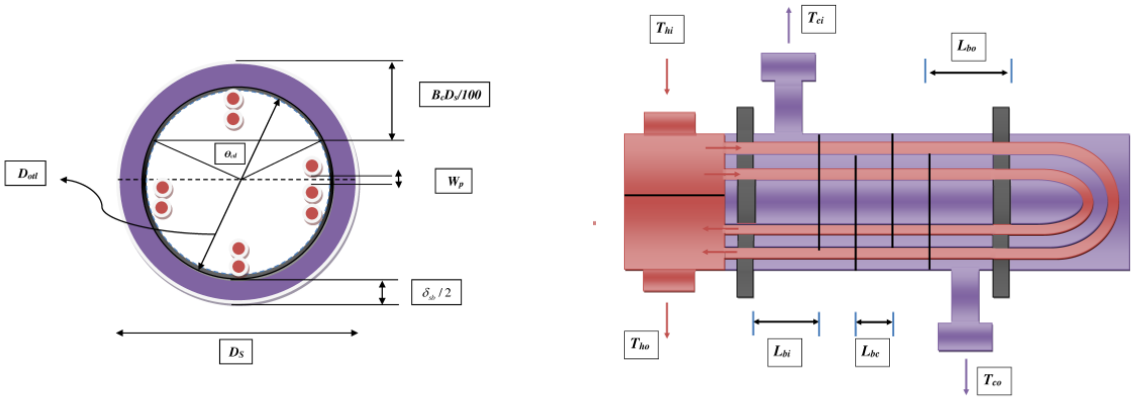

Fig.7. Schematics of a shell-and-tube heat exchanger

$$U_0 = \frac{1}{\frac{1}{h_0} + R_s + \frac{d_o \ln(d_o/d_i)}{2k_W} + R_t \frac{d_o}{d_i} + \frac{1}{h_i}\frac{d_o}{d_i}}, \quad (1)$$

where $h_o$ and $h_i$ are the shell-side and tube-side heat transfer coefficients, $d_o$ and $d_i$ are the outer and inner tube diameters, $R_s$ and $R_t$ are fouling resistances on the shell and tube sides, and $k_w$ is the tube wall thermal conductivity. The heat exchange area is calculated by (2) [2]:

$$A_0 = \frac{Q}{U_0 \Delta T_{lm} F}, \qquad (2)$$

Before calculating the area, the value of $U_0$ should be estimated. After that, the system of equations is solved again using the new calculated area and a corrected area is calculated to be used in the next step. This algorithm continues to a step where the area reaches convergence.

The fundamental equations for heat transfer across a surface is given by [2]:

$$Q = \dot{m}_h c_{p,h}(T_{h,i} - T_{h,o}) = \dot{m}_c c_{p,c}(T_{c,o} - T_{c,i}), \qquad (3)$$

The log-mean temperature difference ($\Delta T_{lm}$) is defined as [2]:

$$\Delta T_{lm} = \frac{(T_{h,i} - T_{c,o}) - (T_{h,o} - T_{c,i})}{\ln\left(\dfrac{T_{h,i} - T_{c,o}}{T_{h,o} - T_{c,i}}\right)}, \qquad (4)$$

$F$ is the log-mean temperature difference correction factor for a special layout. It can be shown that in general it depends on the heat capacity rate ratio ($R$) and correctness coefficient (S) [2]:

$$F \frac{\sqrt{(R^2+1)\ln(\frac{1-S}{1-RS})}}{(R-1)\ln\left[\frac{2-S(R+1-\sqrt{R^2+1})}{2-S(R+1+\sqrt{R^2+1})}\right]} \quad , \qquad (5)$$

$$R = \frac{(T_{h,i} - T_{h,o})}{(T_{c,o} - T_{c,i})}, \qquad (6)$$

$$S = \frac{T_{c,o} - T_{c,i}}{T_{h,i} - T_{c,i}}, \qquad (7)$$

Tube pitch, centre-to-centre distance between tubes, should not be less than 1/25 times the outside diameter. On the other hand, very high values result in an increment of the shell diameter. As a result:

$$P_t = 1.25 d_o, \qquad (8)$$

Having the tube diameter, length and area values; the required number of tubes can be calculated using (9).

$$N_t = \frac{A}{\pi d_o L_t}, \qquad (9)$$

The outer tube diameter limit is calculated as:

$$D_{otl} = d_o \left(\frac{N_t}{K_1}\right)^{\frac{1}{n_1}}, \tag{10}$$

$K_1$ and $n_1$ values for different tube configurations are given in Table 4 [28].

| Number of passes | Triangular pitch | | Square and rotated square | |
|---|---|---|---|---|
| | $K_1$ | $n_1$ | $K_1$ | $n_1$ |
| 1 | 0.319 | 2.142 | 0.215 | 2.207 |
| 2 | 0.249 | 2.207 | 0.156 | 2.291 |
| 3 | 0.175 | 2.285 | 0.158 | 2.263 |

Table 4. Parameters used for the calculation of the tube bundle diameter [22]

The shell diameter is calculated from:

$$D_s = \frac{D_{otl}}{0.95} + \delta_{sb}, \tag{11}$$

Where, $\delta_{sb}$ is the shell-to-baffle clearance.

*Shell-side heat transfer*

The shell-side heat transfer coefficient $h_s$ is determined by Eq. 12 by means of correcting the ideal heat transfer coefficient $h_{id}$ for various leakage and bypass flow streams in a baffled segment of the shell-and-tube exchanger. The $h_{id}$ is determined for pure cross flow in a rectangular tube bank assuming that the entire shell-side stream flows across the tube bank at or near the centreline of the shell. It is computed from the Chilton and Colburn $j$ factor.

Many formulations are recommended for the calculation of these values. The following equation is suggested by Shah and Sekulic [2]:

$$h_s = h_{id} J_c J_\ell J_s J_r, \qquad (12)$$

$$h_{id} = \frac{j \dot{m}_s c_{ps} \Pr^{-\frac{2}{3}}}{A_{o,cr}}, \qquad (13)$$

In the above equation, $A_{o,cr}$ is the flow area at or near the shell centerline for one cross-flow section in the shell-and-tube exchanger, $j$ is the colburn factor, $c_{ps}$ is the specific heat capacity of the fluid in the shell side, where the Prandtl number is given by:

$$\Pr = \frac{c_{ps} \mu_s}{k_s}, \qquad (14)$$

For the calculation of the $j$ parameter value, a set of correlations are used [2, 7]:

$$j = a_1 \left( \frac{1.33}{P_T / d_o} \right)^a (\mathrm{Re}_s)^{a_2}, \qquad (15)$$

$$a = \frac{a_3}{1 + 0.14 (\mathrm{Re}_s)^{a_4}}, \qquad (16)$$

The values of $a$, $a_1$, $a_2$, and $a_3$ are given in Table 5 [28]. Additionally, $Re_s$ is calculated from the following equation.

| Layout angle | Reynolds number | $a_1$ | $a_2$ | $a_3$ | $a_4$ | $b_1$ | $b_2$ | $b_3$ | $b_4$ |
|---|---|---|---|---|---|---|---|---|---|
| **45º** | $10^5$-$10^4$ | 0.370 | -0.396 | 1.930 | 0.500 | 0.303 | -0.126 | 6.59 | 0.520 |
| | $10^4$-$10^3$ | 0.370 | -0.396 | - | - | 0.333 | -0.136 | - | - |
| | $10^1$-$10^2$ | 0.730 | -0.500 | - | - | 3.500 | -0.476 | - | - |
| | $10^2$-$10^1$ | 0.498 | -0.656 | - | - | 26.200 | -0.913 | - | - |
| | <10 | 1.550 | -0.667 | - | - | 32.000 | -1.000 | - | - |

Table 5. Colburn factor (*j*) coefficients and ideal friction factor ($f_{id}$) [28]

$$\text{Re}_s = \frac{\dot{m}_s d_o}{\mu_s A_{o,cr}}, \qquad (17)$$

The correction factor for the baffle configuration ($J_c$) is dependent on the fraction of the total number of tubes in cross flow between baffle tips [2]:

$$J_c = 0.55 + 0.72 F_c, \qquad (18)$$

where $F_c$ represents the fraction of the total number of tubes in the cross-flow section. The $J_c$ value is 1.0 for heat exchangers with no tubes. This values increases to 1.15 for small baffle cuts and decreases to 0.65 for large baffle cuts [2].

The angle in radians between the baffle cut and two radii of a circle through the canters of the outermost tubes is as follows [2]:

$$\theta_{ctl} = 2\cos^{-1}\left(\frac{D_s - 2\ell_c}{D_{ctl}}\right), \qquad (19)$$

$$F_c = 1 - 2F_w = 1 - \frac{\theta_{ctl}}{\pi} + \frac{\sin\theta_{ctl}}{\pi}, \qquad (20)$$

$$D_{ctl} = D_{otl} - d_o. \qquad (21)$$

The correction factor for baffle leakage effects ($J_l$), including both, tube-to-baffle and baffle-to-shell leakages with heavy weight constructions, is given by (22). It is function of the ratio of the total leakage area per baffle to the cross-flow area between adjacent baffles, and also of the ratio of the shell-to-baffle leakage area to tube-to-baffle leakage area. If baffles are too close, $J_l$ will have a lower value, due to higher flows of leakage streams. A typical value of $J_l$ is in the range 0.7 to 0.8.

$$J_\ell = 0.44(1 - r_s) + [1 - 0.44(1 - r_s)]e^{-2.21 r_{lm}}, \qquad (22)$$

$$r_s = \frac{A_{o,sb}}{A_{o,sb} + A_{o,tb}}, \quad r_{lm} = \frac{A_{o,sb} + A_{o,tb}}{A_{o,cr}}, \qquad (23)$$

The correction factor for bundle and pass partition bypass streams ($J_b$) varies from 0.9, for a relatively small clearance between the outermost tubes and the shell for fixed tube sheet construction, to 0.7, for large clearances in pull-though floating head construction. Its value can be increased from 0.7 to 0.9 by proper use of the sealing strips in a pull-through bundle.

$$J_b = \begin{cases} 1 & for N_{ss}^+ \geq 1/2 \\ \exp[-Cr_b(1-(2N_{ss}^+)^{1/3})] & for N_{ss}^+ \leq 1/2 \end{cases}, \quad (24)$$

$$r_b = \frac{A_{o,bp}}{A_{o,cr}}, \quad (25)$$

$$N_{ss}^+ = \frac{N_{ss}}{N_{r,cc}}, \quad (26)$$

$$C = \begin{cases} 1.35 & for \ Re_s \leq 100 \\ 1.25 & for \ Re_s > 100 \end{cases}, \quad (27)$$

The magnitude of the cross-flow area of the flow bypass is given by:

$$A_{o,bp} = L_{b,c}(D_s - D_{otl} + 0.5N_p w_p), \quad (28)$$

The number of tube rows $N_{r,cc}$, crossed through one cross-flow section between baffle tips may be obtained from a drawing or directly counting. It may be also estimated from the following expression:

$$N_{r,cc} = \frac{D_s - 2\ell_c}{X_\ell}, \quad (29)$$

The cross-flow area at or near the shell centreline for one cross-flow section may be estimated from:

$$A_{o,cr} = [D_s - D_{otl} + \frac{D_{ctl}}{X_t}(X_t - d_o)]L_{b,c}, \qquad (30)$$

$J_s$ is the correction factor for larger baffle spacing at the inlet and outlet sections compared to the central baffle spacing. The nozzle locations result in larger end baffle spacing and lower velocities, and thus lower heat transfer coefficients. $J_s$ usually varies from 0.85 to 1.0.

$$J_s = \frac{N_b - 1 + (L_i^+)^{1-n} + (L_o^+)^{1-n}}{N_b - 1 + L_i^+ + L_o^+}, \qquad (31)$$

$$L_i^+ = \frac{L_{b,i}}{L_{b,c}}, \quad L_o^+ = \frac{L_{b,o}}{L_{b,c}}, \qquad (32)$$

$$n = \begin{cases} 0.6 \text{ for turbulent flow} \\ 1/3 \text{ for laminar flow} \end{cases}, \qquad (33)$$

In this study, it was assumed that $L_{b,c} = L_{b,i} = L_{b,o}$, and thus $J_s = 1$. The correction factor for adverse temperature gradient in laminar flow, $J_r$, was not taken into consideration in this study and was set to 1.

*Tube-side heat transfer*

When designing heat exchangers, pressure drop considerations are commonly important factors that must be studied in detail. The heat transfer coefficient of the tube side, $h_i$, is given by [2]:

$$h_i = 0.023 \frac{k_t}{d_i} \Pr_t^{1/3} \operatorname{Re}_t^{0.8} \left(\frac{\mu_t}{\mu_{tw}}\right)^{0.14}, \qquad (34)$$

where the tube-side Reynolds number is:

$$\operatorname{Re}_t = \frac{\rho_t V_t d_i}{\mu_t}, \qquad (35)$$

the Prandl number of the liquid in the tube-side is:

$$\Pr_t = \frac{c_{pt} \mu_t}{\mu_t}, \qquad (36)$$

The velocity of the fluid in the tubes, $V_t$, is calculated with the following equation [2]:

$$V_t = \frac{N_p}{N_t} \frac{\dot{m}_t}{\pi (d_i^2/4) \rho_t}, \qquad (37)$$

*Shell-side pressure drop*

There are several ways to estimate the pressure drop in the shell side. In this work, the shell-side pressure drop is calculated using the Bell-Delaware method, given by the next equation [20]:

$$\Delta P_s = [(N_b - 1)\Delta P_{b,id} \zeta_b + N_b \Delta P_{w,id}]\zeta_l + 2\Delta P_{b,id}\left(1 + \frac{N_{r,cw}}{N_{r,cc}}\right)\zeta_b \zeta_s, \qquad (38)$$

where, $N_{r,cw}$ is the number of effective tube rows in cross flow in each window, and $\Delta P_{b,id}$ is the pressure drop for liquid flowing in an ideal cross flow between two baffles. It is calculated by (39) [20]:

$$\Delta P_{b,id} = \frac{4 f_{id} G_s^2 N_{r,cc}}{2 \rho_s} \left(\frac{\mu_{sw}}{\mu_s}\right)^{0.25}, \qquad (39)$$

The friction factor $f_{id}$ associated with the ideal cross flow is expressed as [20]:

$$f_{id} = b_1 \left(\frac{1.33}{P_T/d_o}\right)^b \mathrm{Re}_s^{b_2}, \tag{40}$$

$$b = \frac{b_3}{1+0.14(\mathrm{Re}_s)^{b_4}}, \tag{41}$$

$$\Delta P_{w,id} = \frac{(2+0.6 N_{r,cw})\dot{m}_s^2}{2\rho_s A_{o,cr} A_{o,w}}, \tag{42}$$

The correction factor $\zeta_b$ is calculated as:

$$\zeta_b = \begin{cases} \exp\left(-2.7 r_b [1-(2N_s^{++})^{1/3}]\right) & \text{for } N_s^{++} < 1/2 \\ 1 & \text{for } N_s^{++} \geq 1/2 \end{cases}, \tag{43}$$

The second correction factor $\zeta_l$ is:

$$\zeta_l = \exp(-1.33(1+r_s) r_{lm}^p), \tag{44}$$

$$p = [-0.15(1+r_s)+0.8], \tag{45}$$

$$\zeta_s = \left(\frac{L_{b,c}}{L_{b,o}}\right)^{1.8} + \left(\frac{L_{b,c}}{L_{b,i}}\right)^{1.8}, \tag{46}$$

*Tube-side pressure drop*

The tube-side pressure drop is calculated from the following expression [20]:

$$\Delta P_t = N_p \left( \frac{4fL}{d_i} + 2.5 \right) \frac{\rho_t V_t^2}{2}, \tag{47}$$

where $f$ is the friction factor for turbulent flow and is given by [20]:

$$f = 0.046 (\text{Re}_t)^{-0.2}, \tag{48}$$

The pumping power for the tube and shell sides is calculated similarly on both sides [20]:

$$P_{s,t} = \frac{\Delta P_t \dot{m}_t}{\rho_t \eta} + \frac{\Delta P_s \dot{m}_s}{\rho_s \eta}, \tag{49}$$

where, $\eta$ is considered to be 0.85 [3].

*Cost analysis*

Cost is always one of the most important factors to take into consideration when designing industrial equipment. Cost can be broken into two principal components, capital cost and operating cost. In addition, maintenance cost incurred during operation is also an important factor that must be estimated; nevertheless they tend to be commonly independent of the size of the heat exchanger. The purchase cost is obtained from the following correlation for ambient operating pressure and carbon steel material [29].

$$\log C_p = K_1 + K_2 \log A_o + K_3 (\log A_o)^2. \qquad (50)$$

$K_1$, $K_2$ and $K_3$ parameters were determined for a shell-and-tube heat exchanger at a particular point in time. As a result, the purchase cost corrected, for the effect of changing economic conditions and inflation, can be calculated with the following correlation.

$$C_2 = C_1 (\frac{I_2}{I_1}) \quad, \qquad (51)$$

where $C$ is the purchase equipment cost, $I$ is the cost index, subscript 1 indicates the base time when the cost was determined and subscript 2 the time when the cost is estimated.

The bare module cost $C_{BM}$ of the heat exchanger, which includes the direct and indirect costs for non-base conditions such as nonambient pressure and materials of construction different from carbon steel, is given by the following correlation:

$$C_{BM} = C_P F_{PM}^o = C_P (B_1 + B_2 F_M F_P), \qquad (52)$$

The pressure factor $F_P$ is given by:

$$\log F_P = C_1 + C_2 \log P + C_3 \log(P)^2, \qquad (53)$$

P is measured in bar gauge, the material factor $F_M$, as well as the $C_1$, $C_2$, and $C_3$ coefficients are listed in Table 6 [29].

| Correlation factor | Value |
| --- | --- |
| $K_1$ | 3.2138 |
| $K_2$ | 0.2688 |
| $K_3$ | 0.07961 |
| $C_1$ | 0 |
| $C_2$ | 0 |
| $C_3$ | 0 |
| $F_M$(shell-CS Tube-Cu) | 1.25 |
| $F_M$ (shell-CS Tube-SS) | 1.7 |
| $B_1$ | 1.8 |
| $B_2$ | 1.5 |

Table 6. Capital cost factors [29]

Pumps must provide work to overcome the pressure drop on the tube side, as well as on the shell side. The annual operating cost, calculated from the total pumping power ($P_{s,t}$) on the tube and shell sides, is given by [29]:

$$OC = 8232 P_{s,t} ec, \qquad (54)$$

where *ec* is the electricity cost. In the previous equation, *ec* is assumed to be $0.1 kW$^{-1}$h$^{-1}$ [29]. The factor 8232 accounts for the number of hours of operation, assuming that the heat exchanger is operating 49 weeks during the year.

The TAC of the heat exchanger is expressed in terms of equal annuities of the bare module cost and the annual operating cost:

$$TC = C_{BM} \frac{i(i+1)^n}{(i+1)^n - 1} + OC, \qquad (55)$$

where *i* is the fractional interest rate per year ($i = 0.05$) and *n* is the expected lifespan of the heat exchanger, which was taken to be 20 years, in order to compare the results obtained in this work with results from the previously published design in Wildi-Tremblay and Gosselin [7].

**Apendix 2**

A Markov chain is a stochastic process that transitions from one state to another using a simple sequential procedure. A Markov chain starts at some state *x (1)* and use a transition function $p(x(t)|x(t-1))$ to determine the next state, *x (2)* conditionally dependent on the previous state, the process must keep iterating to create a sequence of X. In this iterative procedure, the next state of the chain at t+1 is based only on the previous state at *t*. This is an important property when using Markov chains for MCMC because previous states of the chain do not affect the state of the chain after one step.

Markov chains converge to a stationary distribution regardless of the starting point. When this property is applied to MCMC, it allows drawing samples from a distribution using a sequential procedure where the starting state of the sequence does not affect the final estimation process.

*Adaptive metropolis-hastings methods*

The basic idea is to create a Gaussian proposal distribution with a covariance matrix which is calibrated using the sample path of the MCMC chain. The Gaussian proposal is centered at the current position of the Markov chain, $X_n$, and its covariance is given by: $n = s_d Cov(X_0,\ldots,X_{n-1}) + s_d \varepsilon I_d$, where $s_d$ is a parameter that depends only on the dimension $d$ of the state space where $\pi$ is defined. $\varepsilon > 0$ is a constant that we may choose very small. $I_d$ denotes the $d$-dimensional identity matrix [30]. In order to start the adaptation procedure, an arbitrary strictly positive definite initial covariance, $C_0$, is chosen according to a priori knowledge. A time index, $n_0 > 0$, defines the length of the initial non-adaptation period.

$$C_n \begin{cases} C_0 & n \leq n_0 \\ s_d Cov(X_0,\ldots X_{n-1}) + s_d \varepsilon I_d & n > n_0 \end{cases}, \qquad (56)$$

The empirical covariance matrix is determined by the points $X_0,\ldots,X_k \in R^d$:

$$Cov(X_0,\ldots X_{n-1}) = \frac{1}{k}\left(\sum_{i=0}^{k} X_i X_i^T - (k+1)\bar{X}_k \bar{X}_k^T\right), \qquad (57)$$

where $\bar{X}_k = \frac{1}{k+1}\sum_{i=1}^{k} X_i$ and the elements $X_I \in R_d$ are considered as column vectors.

Substituting (56) in (57), it is obtained that, the covariance $C_n$ satisfies the recursive formula when $n > n_0$:

$$C_{n+1} = \frac{n-1}{n}C_n + \frac{sd}{n}\left(n\bar{X}_{n-1}\bar{X}_{n-1}^T - (n+1)\bar{X}_n\bar{X}_n^T + X_n X_n^T + \varepsilon I_d\right) \quad (58)$$

which permits the calculation of the covariance matrix without excessive computational cost, since the mean, $X_n$, also satisfies a recursive formula.

*Delayed rejection*

Delaying Rejection (DR) is a strategy that improves the Metropolis-Hasting (MH) algorithm. Assuming that the position of the chain at time t is $X_t = x$. A candidate $y_1$ is generated from $q_1(x,dy)$ and accepted with the following probability [27].

$$\alpha(x, y_1) = 1 \wedge \frac{\pi(y_1)q_1(y_1, x)}{\pi(x)q_1(x, y_1)} = 1 \wedge \frac{N_1}{D_1}, \quad (59)$$

Upon rejection, instead of retaining the same position, $X_{n+1} = x$, as we would do in a standard MH, a second stage move, $Y_2$, is proposed. The second stage proposal depends not only on the current position of the chain but also on what we have just proposed and rejected: $q2(x, y1, y2)$, the second stage proposal is then accepted with the following probability [27].

$$\alpha_2(x, y_1, y_2) = 1 \wedge \frac{\pi(y_2) q_1(y_2, y_1) q_2(y_2, y_1, x)[1 - \alpha_1(y_2, y_1)]}{\pi(x) q_1(x, y_1) q_2(x, y_1, y_2)[1 - \alpha_1(x, y_1)]}$$
$$= 1 \wedge \frac{N_2}{D_2}$$
, (60)

This process of DR can be iterated. If $q_i$ denotes the proposal at the $i$-th stage, the acceptance probability at that stage is as follows [31]:

$$\alpha_i(x, y_1, ..., y_i) = 1 \wedge \frac{N_i}{D_i},$$
(61)

If the $ith$ stage is reached, it means that $Nj < Dj$ for $j = 1, ..., i - 1$, therefore $\alpha j\ (x, y1, ..., yj)$ is simply $N_j/D_j$, $j = 1, ..., i - 1$ and a recursive formula is obtained.

$$D_i = q_i(x, ..., y_i)(D_{i-1} - N_{i-1}),$$
(62)

which leads to,

$$D_i = q_i(x, ..., y_i)[q_{i-1}(x, ..., y_{i-1})[q_{i-2}(x, ..., y_{i-2})...$$
$$[q_2(x, y_1, y_2)[q_1(x, y_1)\pi(x) - N_1] - N_2] - N_3]...N_{i-1}],$$
(63)

Since all acceptance probabilities are computed in such a way that reversibility with respect to π is preserved separately at each stage, the process of DR can be interrupted at any stage, therefore it can be decided in advancethe number of tries before moving away from the current position. Alternatively, upon each rejection, it can be tossed a p-coin (i.e., a coin with head probability equal to p), and if the outcome is head, it is moved to a higher stage proposal, otherwise it stays put.

**Nomenclature**

| | |
|---|---|
| $\Sigma$ | Variance |
| $A_o$ | Heat transfer surface area (m$^2$) |
| $A_{o,cr}$ | Flow area at or near the shell centreline for one cross-flow section (m$^2$) |
| $A_{o,sb}$ | Shell-to-baffle leakage flow area (m$^2$) |
| $A_{o,tb}$ | Tube-to-shell leakage flow area (m$^2$) |
| $B_c$ | Baffle cut |
| $c_p$ | Specific heat capacity (J kg$^{-1}$ K) |
| $d_i$ | Inside tube diameter (m) |
| $d_o$ | Outside tube diameter (m) |
| $D_{otl}$ | Tube bundle outer diameter (m) |
| $D_s$ | Shell diameter (m) |
| $F$ | Correction factor for number of tube passes, friction factor |
| $G$ | Fluid mass velocity (kg m² s⁻¹) |
| $h$ | Heat transfer coefficient (W m⁻² K⁻¹) |
| $J$ | Correction factor for shell-side heat transfer |
| $k$ | Thermal conductivity (W m⁻¹ K⁻¹) |
| $L_{bc} = L_{bo} = L_{bi}$ | Baffle spacing at center, inlet, and outlet (m) |
| $L$ | Tube length (m) |
| $\dot{m}$ | Mass flow rate (kg s⁻¹) |
| $N_b$ | Number of baffles |
| $N_p$ | Number of tube passes |
| $N_{ss}$ | Number of sealing strip pairs |
| $N_t$ | Total number of tubes |
| $Pr$ | Prandtl number |
| $P_t$ | Tube pitch (m) |
| $P_{s,t}$ | Pumping power on tube and shell sides (W) |
| $Q$ | Heat duty (W) |
| $R$ | Fouling resistance (m² k W⁻¹) |
| *Re* | Reynolds number |
| *T* | Tube thickness (m), Temperature (°C) |
| $U_o$ | Overall heat transfer coefficient (W m$^{-2}$ K$^{-1}$) |
| *v* | Flow velocity (m s$^{-1}$) |

*Greek symbols*

| | |
|---|---|
| $\zeta$ | Shell-side pressure drop correction factor |
| $\mu$ | Viscosity (Pa s) |
| $\delta$ | Density (kg m$^{-3}$) |
| $\delta_{tb}$ | Tube-to-baffle diameter clearance |
| $\delta_{sb}$ | Shell-to-baffle diametrical clearance |
| $\eta$ | Efficiency |
| $\theta_{ctl}$ | Angle in radians |
| $\Delta P$ | Pressure drop |
| $\Delta T_{lm}$ | Log-Mean temperature difference |

*Subscripts and superscripts*

| | |
|---|---|
| *C* | Cold fluid, center of the heat exchanger |
| *H* | Hot fluid |
| *I* | Tube inlet |
| *Id* | Ideal |
| *O* | Tube outlet |
| *S* | Shell side |
| *T* | Tube side |
| *W* | Tube wall |

**References**

[1] S. Kakac, H. Liu, A. Pramuanjaroenkij, Heat exchangers: selection, rating, and thermal design, CRC press, 2012.

[2] R.K. Shah, D.P. Sekulic, Fundamentals of heat exchanger design, John Wiley & Sons, 2003.

[3] S. Fettaka, J. Thibault, Y. Gupta, Design of shell-and-tube heat exchangers using multiobjective optimization, International Journal of Heat and Mass Transfer, 60 (2013) 343-354.

[4] E.U. Schlunder, Heat exchanger design handbook, (1983).

[5] S.T.M. Than, K.A. Lin, M.S. Mon, Heat exchanger design, World Academy of Science, Engineering and Technology, 46 (2008) 604-611.

[6] E. Saunders, Heat exchangers, (1988).

[7] P. Wildi-Tremblay, L. Gosselin, Minimizing shell-and-tube heat exchanger cost with genetic algorithms and considering maintenance, International Journal of Energy Research, 31 (2007) 867-885.

[8] K.J. Bell, Final report of the cooperative research program on shell and tube heat exchangers, University of Delaware, Engineering Experimental Station, 1963.

[9] R. Selbaş, Ö. Kızılkan, M. Reppich, A new design approach for shell-and-tube heat exchangers using genetic algorithms from economic point of view, Chemical Engineering and Processing: Process Intensification, 45 (2006) 268-275.

[10] A. Hadidi, A. Nazari, Design and economic optimization of shell-and-tube heat exchangers using biogeography-based (BBO) algorithm, Applied Thermal Engineering, 51 (2013) 1263-1272.

[11] J.M. Ponce-Ortega, M. Serna-González, A. Jiménez-Gutiérrez, Use of genetic algorithms for the optimal design of shell-and-tube heat exchangers, Applied Thermal Engineering, 29 (2009) 203-209.

[12] M. Fesanghary, E. Damangir, I. Soleimani, Design optimization of shell and tube heat exchangers using global sensitivity analysis and harmony search algorithm, Applied Thermal Engineering, 29 (2009) 1026-1031.

[13] A.C. Caputo, P.M. Pelagagge, P. Salini, Heat exchanger design based on economic optimisation, Applied Thermal Engineering, 28 (2008) 1151-1159.

[14] A. Hadidi, A robust approach for optimal design of plate fin heat exchangers using biogeography based optimization (BBO) algorithm, Applied Energy, 150 (2015) 196-210.

[15] Y. Özçelik, Exergetic optimization of shell and tube heat exchangers using a genetic based algorithm, Applied Thermal Engineering, 27 (2007) 1849-1856.

[16] R. Hilbert, G. Janiga, R. Baron, D. Thévenin, Multi-objective shape optimization of a heat exchanger using parallel genetic algorithms, International Journal of Heat and Mass Transfer, 49 (2006) 2567-2577.

[17] O.E. Turgut, Hybrid Chaotic Quantum behaved Particle Swarm Optimization algorithm for thermal design of plate fin heat exchangers, Applied Mathematical Modelling, 40 (2016) 50-69.

[18] H.V.H. Ayala, P. Keller, M. de Fátima Morais, V.C. Mariani, L. dos Santos Coelho, R.V. Rao, Design of heat exchangers using a novel multiobjective free search differential evolution paradigm, Applied Thermal Engineering, 94 (2016) 170-177.

[19] S. Huang, Z. Ma, F. Wang, A multi-objective design optimization strategy for vertical ground heat exchangers, Energy and Buildings, 87 (2015) 233-242.

[20] D. Habimana, Statistical Optimum Design of Heat Exchangers, (2009).

[21] D. Sorensen, D. Gianola, Likelihood, Bayesian, and MCMC methods in quantitative genetics, Springer Science & Business Media, 2007.

[22] A.F. Smith, G.O. Roberts, Bayesian computation via the Gibbs sampler and related Markov chain Monte Carlo methods, Journal of the Royal Statistical Society. Series B (Methodological), (1993) 3-23.

[23] C. Sherlock, P. Fearnhead, G.O. Roberts, The random walk Metropolis: linking theory and practice through a case study, Statistical Science, (2010) 172-190.

[24] M. Bédard, R. Douc, E. Moulines, Scaling analysis of delayed rejection MCMC methods, Methodology and Computing in Applied Probability, 16 (2014) 811-838.

[25] H. Perry Robert, W. Green Don, O. Maloney James, Perry's chemical engineers' handbook, Mc Graw-Hills New York, (1997) 56-64.

[26] V.J. Stachura, Standards of the tubular exchanger manufacturers association, Tubular Exchanger Manufacturers Association, 1988.

[27] H. Haario, M. Laine, A. Mira, E. Saksman, DRAM: efficient adaptive MCMC, Statistics and Computing, 16 (2006) 339-354.

[28] G. Towler, R.K. Sinnott, Chemical engineering design: principles, practice and economics of plant and process design, Elsevier, 2012.

[29] R. Turton, R.C. Bailie, W.B. Whiting, J.A. Shaeiwitz, Analysis, synthesis and design of chemical processes, Pearson Education, 2008.

[30] H. Haario, E. Saksman, J. Tamminen, An adaptive Metropolis algorithm, Bernoulli, 7 (2001) 223-242.

[31] A. Mira, On Metropolis-Hastings algorithms with delayed rejection, Metron, 59 (2001) 231-241.